# Radiation damage studies of silicon photomultipliers for the CMS MIP Timing Detector

**Y. Musienko**[a,b,1], **A. Heering**[a], **A. Karneyeu**[a] **and M. Wayne**[a]

[a] *University of Notre Dame, Notre Dame, IN, U.S.*

[b] *Institute of Nuclear Research, Moscow, Russia*

*E-mail*: Iouri.Musienko@cern.ch

ABSTRACT: Recently developed 15 um, 20 um, 25 um and 30 um cell size Hamamatsu Silicon Photomultipliers (SiPMs) were irradiated with reactor neutrons at JSI (Ljubljana) up to $2\times10^{14}$ n/cm$^2$ (1 MeV equivalent). The parameters of new and irradiated SiPMs were studied using pulsed light illumination. The effects of the neutron radiation on breakdown voltage, signal amplitude, dark current and noise for these devices are shown and discussed

[1] Corresponding author.

## Contents

## 1. Introduction

The barrel section of the novel MIP Timing Detector (MTD) [1] will be constructed as part of the upgrade of the CMS experiment to provide a time resolution for single charged tracks in the range of 30 − 60 ps using LYSO:Ce crystal arrays read out with Silicon Photomultipliers (SiPMs). A major challenge for the operation of such a detector is the extremely high radiation level, of about $2 \times 10^{14}$ 1 MeV(Si) Eqv. n/cm$^2$, that will be integrated over a decade of operation of the High Luminosity Large Hadron Collider (HL-LHC). Changes to the SiPM parameters during irradiation is one of the most important subjects regarding the application of SiPMs in high energy physics experiments [2]. Silicon Photomultipliers exposed to high levels of radiation have shown a strong increase in dark count rate and radiation damage effects that also impact their gain and photon detection efficiency [3].

In this article we present the results of neutron irradiation of the newly developed large dynamic range SiPMs from Hamamatsu (Japan).

## 2. SiPMs under study, measurement set-up and irradiation facility.

Large dynamic range SiPMs were developed in 2020-2023 by the CMS SiPM group in cooperation with Hamamatsu for the CMS MTD project. The SiPMs have a 9 mm$^2$ area and 15 μm, 20 μm, 25 μm and 30 μm cell pitch. All SiPMs are of the so-called PIT type (improved PDE in comparison to the Hamamatsu S13360 series SiPMs). In Hamamatsu, the SiPMs were housed inside a ceramic case and protected by silicone rubber (0.30 ± 0.05 mm thickness, index of refraction – 1.57).

A Keithley-6487 picoammeter/voltage source was used to apply bias voltage to the SiPMs and for current measurements. Dark currents vs. overvoltage (V-VB) dependences for the Hamamatsu SiPMs measured at 23 °C are shown in Fig.1a. Fig. 1b shows the dependence of photo detection efficiencies (PDEs) vs. overvoltage. The SiPMs were illuminated with small LED (410 nm average emission wavelength) pulses via a 2 mm collimator. The signals from the SiPMs were amplified with a fast transimpedance amplifier (gain=20) and digitized with a Picoscope 6404D digital oscilloscope. The number of photons in the LED pulse was measured using a calibrated XP2020 photomultiplier (QE(410 nm)=25 %). The SiPMs gains and excess noise factors in dependence on the overvoltage are shown on Fig.1c and Fig.1d. The signal integration time was 150 ns. A Labview based code was written to analyse the recorded waveforms. The setup and the experimental technique used in these measurements are described in detail in [4].

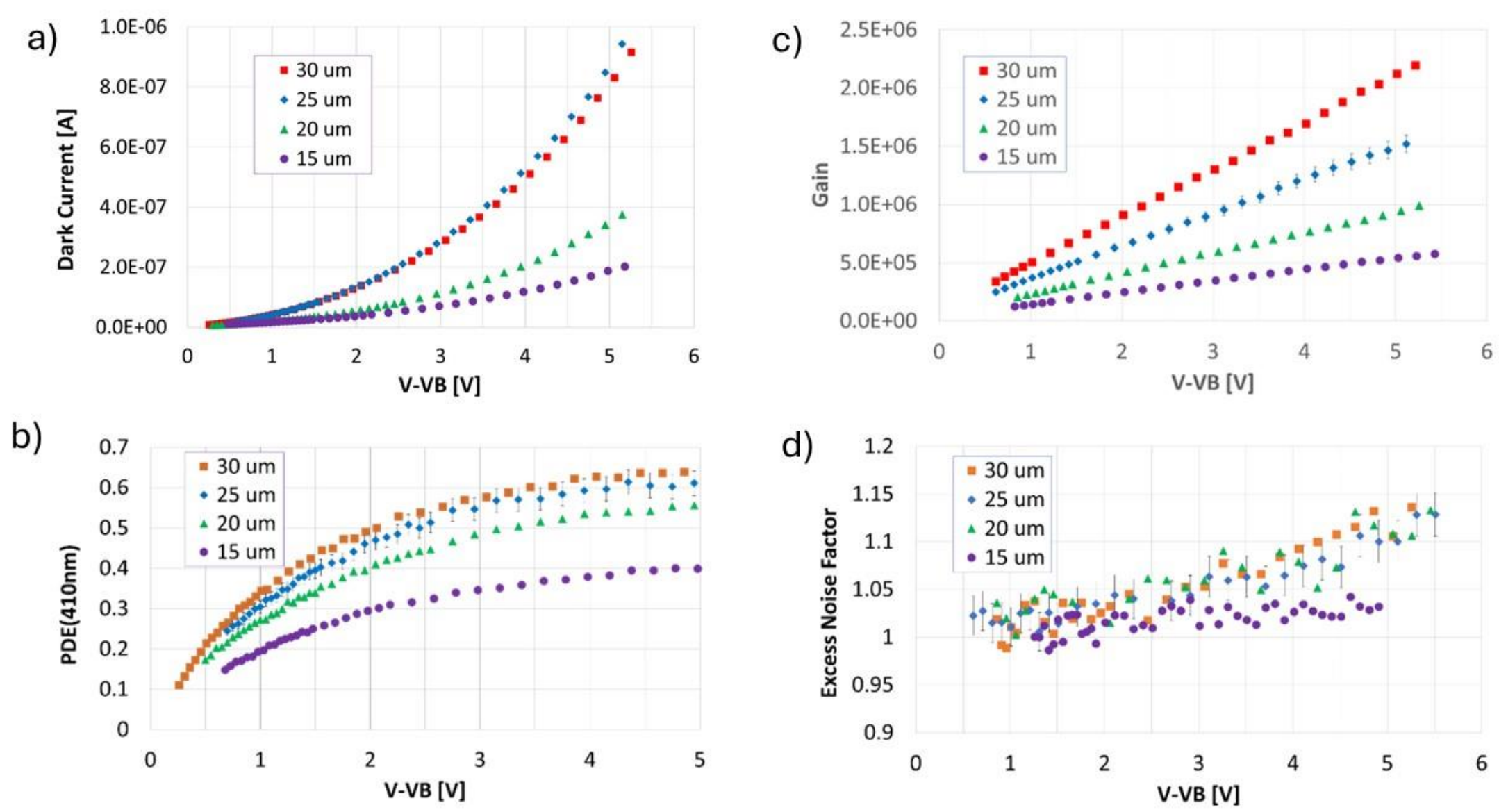

**Figure 1.** Dependence of the most important SiPM parameters on the overvoltage.

The most important parameters of these SiPMs are summarized in Table 1. The SiPM parameters in the table were measured at T=23 °C and dVB=3 V. SiPM breakdown voltages were determined using the maximum relative derivative method described in [7]. An HP4284 LCR meter operated at a frequency of 1 MHz was used to measure the SiPMs capacitances. The cell recovery time was estimated from the SiPM capacitance measurements and forward current measurements.

| Parameter | 15 µm SiPM | 20 µm SiPM | 25 µm SiPM | 30 µm SiPM |
|---|---|---|---|---|
| Cell Pitch, µm | 15 | 20 | 25 | 30 |
| Sensitive Area, mm$^2$ | 9 | 9 | 9 | 9 |
| Breakdown Voltage, V | 38.32 | 38.45 | 38.45 | 38.94 |
| Operating Voltage, V | 41.32 | 41.45 | 41.45 | 41.94 |
| Dark Current, nA | 75 | 120 | 280 | 280 |
| PDE (410 nm), % | 34 | 49 | 55 | 57 |
| Gain, x10$^3$ | 345 | 590 | 910 | 1300 |
| Capacitance, pF | 596 | 624 | 637 | 620 |
| Cell Recovery Time, ns | ~8 | ~10 | ~17 | ~24 |
| Excess noise factor | ~1.03 | ~1.05 | ~1.05 | ~1.05 |
| After-pulse probability, % | <2 | <2 | <2 | <2 |
| dVB/dT, mV/°C | 37 | 37 | 37 | 37 |

**Table 1.** Some of the SiPM parameters before irradiation (T=23 °C, dVB=3 V).

The SiPMs (5 of each cell size) were passively irradiated to a total dose of $2x10^{14}$ n/cm$^2$ (1 MeV equivalent) at the TRIGA reactor at JSI (Jožef Stefan Institute) in Slovenia. "Clones" of the SiPMs (SiPMs produced from the same wafers) were selected and kept non-irradiated for comparative measurements. Their parameters (breakdown voltage, Gain, PDE etc.) were identical to the parameters of the SiPMs which were used for irradiation. After irradiation the diodes were annealed 3.5 days at 110 °C to stabilize their currents [5, 15].

## 3. Experimental results on irradiated SiPMs

The SiPM parameters were measured at the CERN APD Lab: dark current vs. bias, noise vs. bias, S/N ratio vs. bias for constant LED signal. All the measurements were performed at T=-45 °C inside a commercial freezer [6]. The temperature inside freezer was -38 °C. A thermoelectric cooler (TEC) was used to further reduce and stabilize the temperature of the SiPMs. Thermal contact between the thermoelectric cooler and the SiPM ceramic body was ensured using thermally conductive grease.

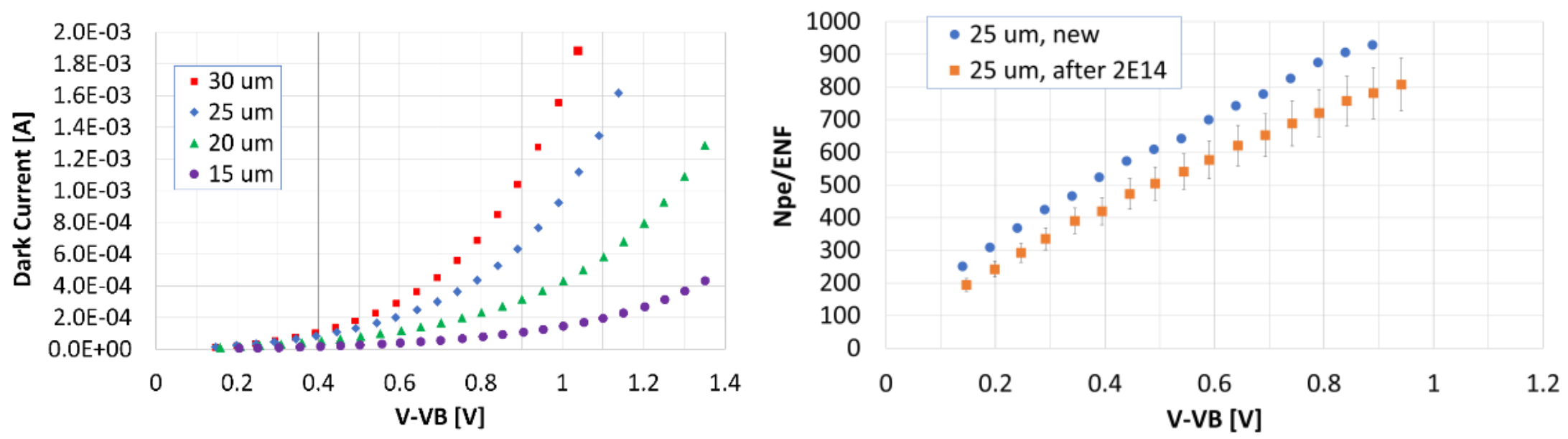

**Figure 2.** Dark current vs. overvoltage measured for the SiPMs after irradiation and annealing at T=-45 °C (left). Dependence of the signal amplitude in photoelectrons on the bias voltage measured for new and irradiated 25 µm SiPMs at T=-45 °C (right).

An increase of breakdown voltages was measured after irradiation for all irradiated SiPMs (1.89 V for the 15 µm SiPM and 0.98 V for the 20 µm, 25 µm and 30 µm SiPMs). The maximum relative derivative method [7] was used to calculate the SiPM breakdown voltages before/after irradiation.

A significant increase in the dark current was measured for all the SiPMs. Figure 2 (left) shows the dependence of the dark current vs. overvoltage measured for four SiPMs after irradiation. Dark currents as high as 150 or 1600 µA (for the 15 and 30 µm SiPMs respectively) were measured at 1 V overvoltage at T=-45 °C after annealing. Such high currents can cause local increase of the SiPM temperature and can be one of the reasons for the measured gain reduction after irradiation [16].

Responses of the SiPMs to calibrated LED pulses as functions of bias voltage were measured for irradiated and non-irradiated SiPMs. The LED (λ=410 nm) was fired by a PM 5786 pulse generator. The amplitude of the LED pulse was the same (~3 300 photons/pulse and ~20 ns in duration) for all the SiPMs (irradiated and non-irradiated). The SiPMs were illuminated via a collimator of 2 mm in diameter. The signals from the SiPMs were amplified with a fast transimpedance amplifier (gain=20) and digitized with a Picoscope 6404D digital oscilloscope. The recorded waveforms were integrated during a 50 ns gate. The results for all the SiPMs are shown in Fig. 3. One can see that the reduction of the measured LED signal for the irradiated SiPM is in the range 25% ÷ 35 % at dVB~1 V. Most of this change is due to the SiPM PDE reduction. The number of photoelectrons/ENF ratio in dependence on overvoltage (shown in Fig. 2 (right)) was calculated using width of the signal amplitude spectra (see details of the method in [2, 8]).

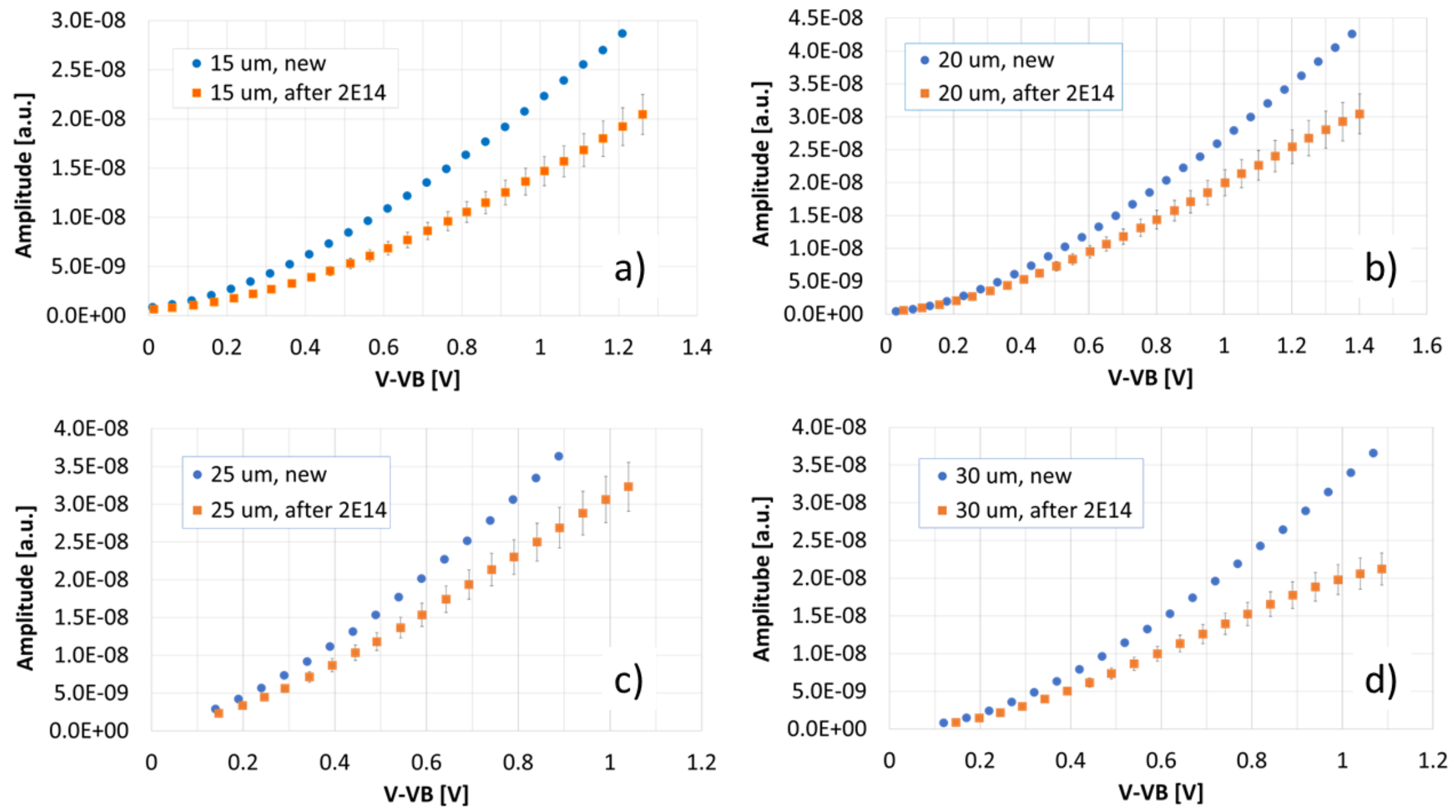

**Figure 3.** Dependence of the signal amplitude on the bias voltage measured for new and irradiated 25 μm SiPMs at T=-45 °C.

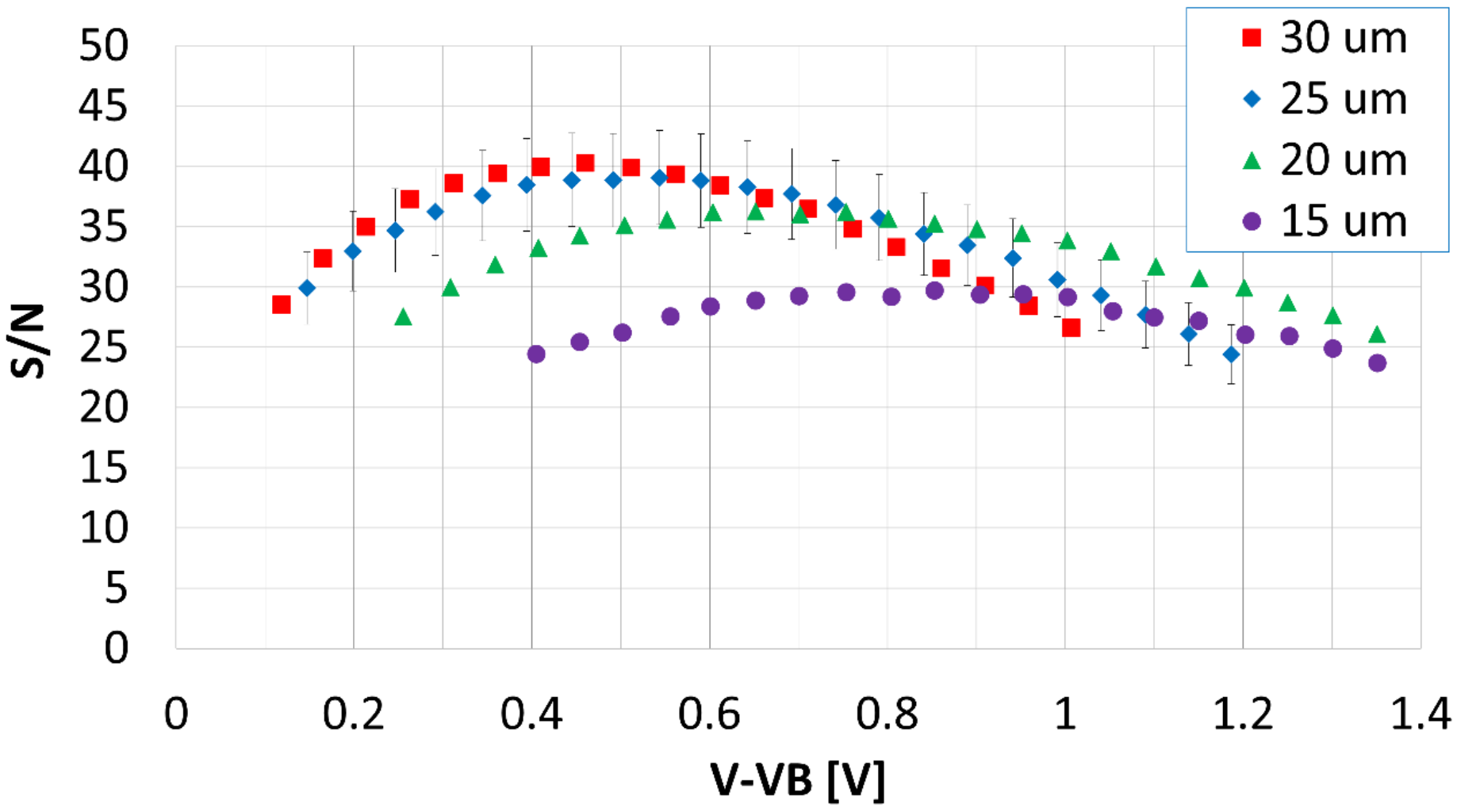

**Figure 4.** Signal to noise ratio vs. overvoltage measured for the SiPMs after irradiation

The calculated signal to noise ratio (ratio of the signal amplitude from the LED to the standard deviation of the pedestal value) dependence on the overvoltage for the irradiated devices is shown in Fig. 4. We see that the signal-to-noise ratio reaches its maximum at rather low SiPM overvoltages (0.4 ÷ 0.9 V). The signal to noise ratio is higher for the 25 um ÷ 30 um Hamamatsu SiPMs due to the higher PDE of these SiPMs.

## 4. Dependence of the dark current of irradiated SiPM on temperature.

The change of dark current in silicon detectors with particle fluence has been extensively studied, see for instance [9] and references therein. The major reason for its increase is the generation of electron-hole pairs due to defects created by irradiation in the depletion region [10], which in the case of SiPMs can be enhanced by a high electric field leading to the mechanism of trap assisted tunnelling [11]. In the case of low electric fields, the generation currents have the following temperature dependence [12]:

$$I_{gen} \sim T^2 e^{-\frac{E_a}{KT}} \qquad (1)$$

For the activation energy, $E_a$ a value of 0.605 eV is found by Chilingarov [13]. For electric fields of the order of $10^5$ V/cm or higher Eq. (1) needs to be corrected by a trap-assisted tunneling term, $I_{gen+tat}$. The correction depends on the effective field strength, $F_{eff}$ and modifies Eq. (1), as [2]:

$$I_{gen+tat} \propto (1+\Gamma)T^2 e^{-\frac{E_a}{KT}} \qquad (2)$$

where $\Gamma = 2\sqrt{3\pi}\frac{F_{eff}}{(KT)^{3/2}}e^{\left(\frac{F_{eff}}{(KT)^{3/2}}\right)^2}$ is the term defined by Hurkx [14], which accounts for the effects of tunneling.

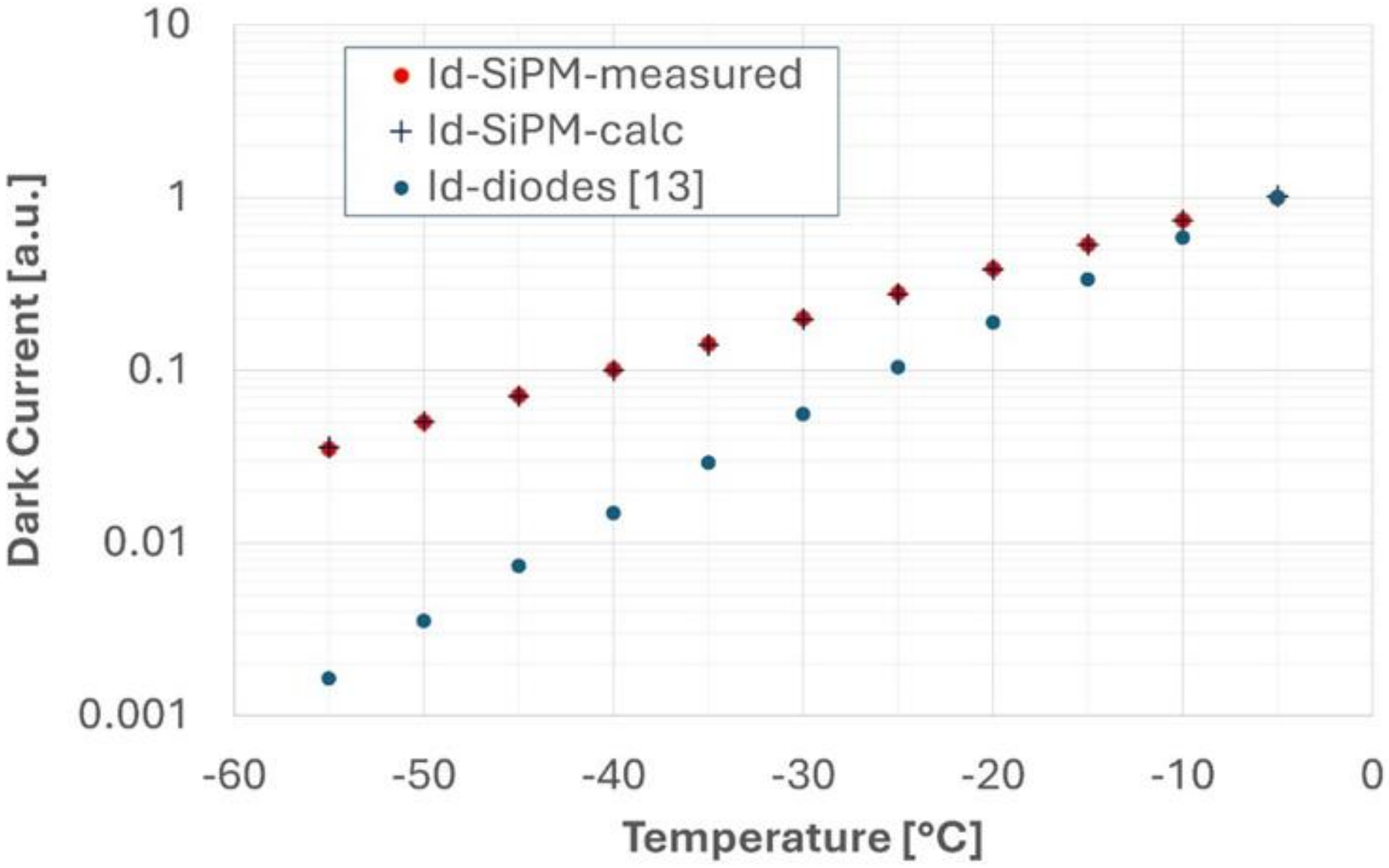

**Figure 5.** Normalized (to the value at T=-5 °C) dark currents vs. temperature for the irradiated 25 µm SiPM and the Si diode [13], measured and calculated.

Dependence of dark current on temperature was studied for the Hamamatsu 25 um cell pitch SiPM irradiated with $5\times10^{13}$ n/cm$^2$ (1 MeV equivalent) and annealed 4 days at 120 °. The set-up with SiPM was placed inside a freezer operated at -40 °C. The temperature of the SiPM was varied using a thermoelectric cooler. The dark current vs. bias voltage was measured at 11 different temperatures (from -5 to -55 °C). Fig. 5 shows the normalized (to the value found at -5 °C) dark currents of the irradiated Hamamatsu SiPM measured at 1 V overvoltage and normalized dark currents of the irradiated silicon sensors from [13] in dependence on temperature. From the results presented in Figure 5, it is evident that the dark current of the irradiated SiPM decreases with decreasing temperature much more slowly than the current of the irradiated silicon diode.

This figure also shows the results of fitting the SiPM data using equation (2) with an activation energy of $E_a$=0.605 eV. The term $F_{eff}$ accounting for trap-assisted tunnelling effects, was found to be $6.3\times10^{-3}$ eV$^{3/2}$.

## 5. Summary

Results from a study of the radiation hardness of silicon photomultipliers (SiPMs) are presented. Recently developed 15 um, 20 um, 25 um and 30 um cell SiPMs from Hamamatsu (Japan) were exposed to reactor neutrons at fluences up to $2\times10^{14}$ neutrons/cm$^2$ (1 MeV equivalent) at JSI (Slovenia). After irradiation the diodes were annealed (3.5 days at 110 °C) to stabilize their currents. Responses of the SiPMs to calibrated LED pulses as functions of bias voltage were measured for irradiated and non-irradiated SiPMs at T=-45 °C. Here we present the main results of our study:

- The SiPMs remain functional after such a high neutron dose.
- We observed a breakdown voltage increase for all SiPMs (1.89 V for 15 um SiPM and 0.98 V for 20 um, 25 um and 30 um SiPMs).
- A significant increase in dark current and noise was measured.
- The LED signal amplitude measured by the irradiated SiPM was about 25% ÷ 35% smaller than that measured by the non-irradiated SiPMs. Most of the reduction is due to a reduction in PDE.
- A signal to noise ratio reaches maximum at 0.4 ÷ 0.9 Volt overvoltage for all SiPMs.

We also studied the temperature dependence of dark current of the irradiated SiPMs. We found that the change in dark current of an irradiated SiPM with decreasing temperature is significantly weaker than that of an irradiated PIN diode. The effect can be related to the difference in electric field strength inside these devices.

The results obtained by this study provided important information for the implementation of SiPMs in the readout of the BTL detector at the LHC, including the choice of 25 μm as the optimal pixel size [17].

## Acknowledgments

The authors would like to express their gratitude to the Jožef Stefan Institute irradiation facility personel for their help with irradiations. This project has received funding from the European Union’s Horizon Europe Research and Innovation programme under Grant Agreement No 101057511 (EURO-LABS). We also thank the Department of Energy and the National Science Foundation (USA) for their support of this work.

## References

[1] The CMS Collaboration, A MIP Timing Detector for the CMS Phase-2 Upgrade, [CERN-LHCC-2019-003; CMS-TDR-020].

[2] E. Garutti and Y. Musienko, Radiation damage of SiPMs, Nucl. Instr. and Meth. A926 (2019) 69

[3] Y. Musienko et al., Effects of very high radiation on SiPMs, Nucl. Instr. and Meth. A 610 (2009) 87.

[4] A. Vacheret et al., Characterization and Simulation of the Response of Multi Pixel Photon Counters to Low Light Levels, Nucl. Instr. and Meth. A 656 (2011) 69.

[5] F. Addesa et al., Optimization of LYSO crystals and SiPM paramteters for the CMS MIP timing detector, 2024 JINST 19 P12020

[6] GFL laboratory Chest freezer 6340, https://profilab24.com/en/laboratory/refrigerators-cooling-technology/gfl-chest-freezer-6340

[7] Y. Musienko, "Radiation damage studies of silicon photomultipliers for the CMS HCAL phase I upgrade", talk given at 7-th NDIP conference, Tours, France, July 2014 (https://ndip2017.event-vert.com/past_editions/tours14/AGENDA/AGENDA-by-DAY/Presentations/5Friday/PM/ID37166_Musienko.pdf)

[8] Y. Musienko et al., Radiation damage in silicon photomultipliers exposed to neutron radiation, 2017 JINST **12** C07030

[9] M. Moll, Radiation Damage in Silicon Particle Detectors, Ph.D. Thesis, Hamburg University, 1999, DESYTHESIS-1999-040, ISSN-1435-8085, (http://mmoll.web.cern.ch/mmoll/thesis/)

[10] G. Lutz, Semiconductor Radiation Detectors - Device Physics, Springer Verlag, Berlin/Heidelberg, ISBN: 978-3540716785, 1999.

[11] G. Vincent, A. Chantre, D. Bois, Electric field effect on the thermal emission of traps in semiconductor junctions, Appl. Phys. 50 (8) (1979) 5484–5487.

[12] A. Grove, Physics and Technology of Semiconductor Devices, John Wiley & Sons, ISBN: 978-0471329985, 1967.

[13] A. Chilingarov, Temperature dependence of the current generated in Si bulk, JINST 8 (10) P10003.

[14] G. Hurkx, D. Klaassen, M. Knuvers, A new recombination model for device simulation including tunneling, IEEE Trans. Nucl. Sci. 39 (2) (1992) 331–338

[15] A. Ledovskoy et al., SiPM annealing model (article in preparation)

[16] M. Lucchini, Y. Musienko, A. Heering, Experimental method to monitor temperature stability of SiPMs operating in conditions of extremely high dark count rate, Nucl. Instr. and Meth. A 977 (2020) 164300

[17] M. Malberti, Precision timing with the CMS Barrel Timing Layer, talk given at PD2025 - 7th international workshop on new Photon-Detectors Bologna 3-5 December 2025, https://agenda.infn.it/event/47073/contributions/276185/